\documentclass[conference]{IEEEtran}
\usepackage[dvips]{graphicx}
\graphicspath{{figs/}}
\usepackage[cmex10]{amsmath}
\usepackage{url}
\usepackage{multirow}
\usepackage{array}
\usepackage{footnote}

\begin{document}
\title{Virtual Private Overlays: Secure Group Communication in NAT-Constrained Environments}

\author{\IEEEauthorblockN{David Isaac Wolinsky, Kyungyong Lee, Tae Woong Choi, P. Oscar Boykin, and Renato Figueiredo} 
\IEEEauthorblockA{Advanced Computing Information Systems Lab\\
University of Florida}
}

\maketitle

\begin{abstract}
Structured P2P overlays provide a framework for building distributed applications that
are self-configuring, scalable, and resilient to node failures.  Such systems
have been successfully adopted in large-scale Internet services such as content
delivery networks and file sharing; however, widespread adoption in small/medium
scales has been limited due in part to security concerns and difficulty
bootstrapping in NAT-constrained environments.  Nonetheless, P2P systems can be designed to
provide guaranteed lookup times, NAT traversal, point-to-point overlay
security, and distributed data stores. In this paper we propose
a novel way of creating overlays that are both secure and
private and a method to bootstrap them using a public overlay.
Private overlay nodes use the public overlay's distributed data store to discover
each other, and the public overlay's connections to assist with NAT hole punching
and as relays providing STUN and TURN NAT traversal techniques.  The security
framework utilizes groups, which are created and managed by users through
a web based user interface.
Each group acts as a Public Key Infrastructure (PKI) relying on
the use of a centrally-managed web site providing an automated Certificate
Authority (CA).  We present a reference implementation which has been used in
a P2P VPN (Virtual Private Network).  To evaluate our contributions, we apply our
techniques to an overlay network modeler, event-driven simulations using simulated
time delays, and deployment in the PlanetLab wide-area testbed.
\end{abstract}
\section{Introduction}
While Structured P2P overlays provide a scalable, resilient, and self-managing
platform for distributed applications, their adoption rate has been slow outside
data distribution services such as BitTorrent
and eDonkey.  General use of structured P2P systems, especially in applications
targeting homes and small/medium businesses (SMBs), has been limited in large
part due to the difficult nature of securing such systems to the level required
by these users.  Applications in home
and SMBs may need a greater level of trust than what can be guaranteed by
anonymous contributors in free-to-join overlays, but these users lack the
resources for bootstrapping private P2P overlays particularly in constrained
wide-area network environments where a significant amount of or all
peers are behind Network Address Translation devices (NAT). This paper presents
a novel approach that enables virtual private overlays 
to be created and managed by members of small/medium groups, leveraging public
overlays for bootstrapping, NAT traversal and relaying.

There are many different P2P applications used in home and small business,
primarily for collaboration and sharing, including data storage, media
sharing, chat, and system maintenance and monitoring.
Applications that currently provide these functionalities fall into two
categories:  anonymous, fully decentralized free-to-join P2P systems and
distributed systems with P2P communication that rely on a third party to provide
discovery and management.  While using a third party service provides a desirable
level of trust for many users, it has significant drawbacks such as vendor lock-in,
which may result in lost data, down time, and scalability constraints.

An example of a useful small business application that falls into the latter
category is LogMeIn's~\cite{logmein} software products LogMeIn Pro and Hamachi.
LogMeIn Pro allows users to remotely manage and connect with their machines so
long as they are willing to use LogMeIn's software and infrastructure.  Hamachi
allows users to establish decentralized VPN links using centralized session
management.  Both applications assist in the remote maintenance and monitoring
of computers without requiring the user to implement the networking
infrastructure provided by LogMeIn.

Some examples of P2P applications for homes and small
businesses in development are P2PSIP and P2Pns.  P2PSIP enables users to initiate, through decentralized means, visual
and audio communication, while P2Pns allows users to deploy a decentralized
naming service.  Both applications allow users to contribute and benefit from
all members of the system without the regulation of a third party, but lack the
ability to allow users to centrally secure and manage their own subset of the
systems.

Distributed data store applications like Dynamo~\cite{dynamo} and
BigTable~\cite{bigtable} have the ability to store data using a completely
decentralized system.  Though these systems are highly scalable and fault
tolerant, the software uses an untrusted overlay.  Therefore all instances
need to run in a secure environment, whether that is in a single institution
or across a largely distributed environment using a VPN.

In this paper, we describe the architecture of a system that attempts
to balance the benefits of third party services with P2P infrastructures.
The main contribution of this paper is the architecture of a P2P messaging
framework that integrates existing datagram-based security
(Section~\ref{secure_overlays}), supports bootstrapping multiple virtual
private overlays using a public overlay to assist in NAT traversal
(Section~\ref{private_overlays}), and binds privacy and security together via
a group infrastructure (Section~\ref{group_overlays}).
Our system relies on a completely open infrastructure using mechanisms that
reduce the maintenance and deployment burden that exist with decentralized
P2P systems without the loss of ownership due to use of third party systems.
The components of our system follow:
\begin{enumerate}
\setlength{\itemsep}{0pt}
\setlength{\parskip}{0pt}
\item An application layer security system based upon an existing datagram-based
transport security (DTLS) framework to provide secure end-to-end (EtE) and
point-to-point (PtP) in P2P overlays\footnote{For our discussion a point-to-point
or PtP relationship refers to direct connections between two peers, such as a UDP
or TCP socket; whereas, an end-to-end or EtE relationship refers to messages passed
through the overlay, which may be routed over many PtP links.}
 while transparently handling security
concerns of relay NAT traversal.
\item The bootstrapping of private P2P overlays using existing public overlays.
The term "public overlay" refers to a free-to-join bootstrap overlay; nodes in
the public overlay are not required to have public IP addresses and can be
behind NATs.
\item A group web interface providing an automated front-end handler for a 
Public Key Infrastructure (PKI) that allows users to create their own secure
systems relying on P2P overlays.
\end{enumerate}

\begin{figure}[h]
\centering
\includegraphics[width=3.25in]{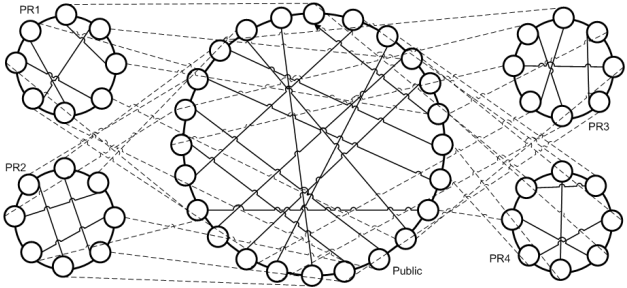}
\caption{The use of a single, public overlay to bootstrap multiple, isolated
private overlays.  The ring in the center is the public overlay.  Each node in
the private overlay has a corresponding partner in the public pool; the
relationship is represented by a dashed line.  A public overlay node can be
multiplexed by more than one private overlay.}
\label{fig:subrings}
\end{figure}

Our approach provides an easy mechanism to create trusted overlays in
constrained environments using a publicly available untrusted overlay as
illustrated in Figure~\ref{fig:subrings}.  Akin to other virtualization
techniques, the virtual private overlay model allows developers to focus
on the application, while complexities associated with enforcing isolation
and security and messaging are abstracted away.
This approach also has benefits for system deployers, whereas insecure systems
would require modifying the overlay software to support security, application
level security, or the deployment of a decentralized and scalable VPN in
addition to the overlay software.  Note, that the usefulness of the public
overlay could be nullified if it does not employ some form of decentralized
security as described in~\cite{secure_routing}.

The rest of this paper is organized as follows.  Section~\ref{structured_p2p}
provides background and related work.
Section~\ref{contributions} describes our contributions.  In
Section~\ref{applications}, we present different usage models and present a
real system using the techniques described in this paper, the GroupVPN.  In
Section~\ref{evaluations}, we described technical details of our implementation
and use models, simulators, and real systems to verify the utility of our
approach.  We conclude the paper in Section~\ref{conclusions}.

\section{Background}
\label{background}
In this section, we begin by reviewing structured P2P overlays, followed by
constraints that make the creation of secure and private P2P overlays
difficult, and related work.

\subsection{Structured P2P Systems}
\label{structured_p2p}
Structured P2P systems provide distributed look up services with guaranteed
search time with a lower bound of $O(\log N)$, in contrast to unstructured
systems, which rely on global knowledge/broadcasts, or stochastic techniques
such as random walks~\cite{unstructured_v_structured}.  Some examples of
structured systems can be found in~\cite{pastry, chord, symphony, kademlia,
can}.  In general, structured systems are able to make these guarantees by
self-organizing a structured topology, such as a 2D ring or a hypercube.

Each node is given a unique node ID.  The node ID, drawn from a large address space, must be unique to each peer,
otherwise address collisions will occur which can prevent nodes from participating
in the overlay.  Furthermore, having the node IDs well distributed assist in
providing better scalability as many algorithms for selection of shortcuts
depend on having node IDs uniformly distributed across the entire address space.
A simple mechanism to ensure this is to have each node use a cryptographically
strong random number generator.  Another mechanism for distributing node IDs
involves the use of a trusted third party to generate node IDs and
cryptographically sign them~\cite{secure_routing}.

As with unstructured P2P systems, in order for an incoming node to connect with
the system it must know of at least one active participant.  A list of nodes
that are running on public addresses should be maintained and distributed with
the application, available through some out-of-band mechanism, or possibly using
multicast to find pools~\cite{pastry}.

Depending on the protocol, a node must be connected to closest
neighbors in the node ID address space; optimizations for fault tolerance suggest
that it should be between 2 to $\log(N)$ on both sides.  
Having multiple peers on both sides assist in stabilizing the overlay structure
when experiencing churn, particularly when peers leave without warning.

Overlay shortcuts enable efficient routing in ring-structured P2P systems.  The
different shortcut selection methods include: maintaining large tables without
using connections and only verifying usability when routing
messages~\cite{pastry, kademlia}, maintaining a connection with a peer every
set distance in the P2P address space~\cite{chord}, or using locations drawn
from a harmonic distribution in the node address space~\cite{symphony}.

Most structured P2P overlays support decentralized storage/lookup of information by
mapping keys to specific node IDs in an overlay.  At a minimum, the data is stored
at the node ID either smaller or larger to the data's node ID and for fault
tolerance the data can be stored at other nodes.  This sort of mapping
and data storage is called a distributed hash table (DHT) and is a typical
component of most structured P2P systems.

\subsection{Constraints in Structured P2P Systems}
\subsubsection{Communication Between Nodes}
There are two mechanisms for message routing in a P2P overlay: iterative or
recursive.  In iterative routing, the sender of a packet will contact each
successive member in a path directly until it find the destination node, at
which point it sends the packet directly to the destination.  In recursive
routing, messages are sent through the overlay via forwarding from one peer to
the next until arriving at the destination.  Iterative routing makes NAT
traversal complicated as peers create connections during the routing of packets,
which would require constant NAT traversal mediation, whereas recursive routing
provides stable connections due to a single NAT traversal during the connection
phase.  In regards to security, Iterative routing can easily be secured using
stream and datagram based security such as TLS~\cite{tls} and DTLS~\cite{dtls},
because the sender initiates all messages; however, in
recursive routing, EtE communication cannot be secured in the same fashion by
through socket-based TLS and DTLS, the sender because messages are routed
through intermediate nodes.

\subsubsection{Private and Secure Overlay Subsets}
Though EtE authentication and privacy is important for many applications, it
does not increase the reliability of the structured
overlay.  In a free-to-join system, malicious peers can easily intercept packets
for eavesdropping or to discard or tamper with them.  To deal with
this issue, each overlay node could participate with a subset of other nodes in
a secure system where each member would maintain a routing table containing a subset
of those involved.  Each node would have to distinguish between the different
messages, handing them to unique routers for each group, handle connectivity
of the subset during churn, and providing a distributed data store for use by this subset.
Achieving this functionality can require substantial modifications to the core overlay
messaging and storage primitives. In this paper we advocate a virtualization approach
that multiplexes private overlays that provide the same abstraction of the underlying
public overlay, and thus can reuse core overlay primitives without modifications.

\subsubsection{NATs}
Another key issue faced by P2P systems is the handling of NATs.
A recent study~\cite{p2p_nats} has shown that 30\%-40\% of nodes in P2P systems
are behind NATs.  In environments where there are NATs, iterative routing can be
significantly more difficult to deploy, since each message sent may require
multiple NAT traversals.  There are a few types of NATs that can be traversed
to support bidirectional communication through UDP-hole punching, known as STUN.
The remaining categories of NATs either do not support UDP-hole punching due to
how the NATs do port mapping or support only TCP communication and block all
UDP communication.  In~\cite{tcp_nat, p2p_nat}, the authors evaluated many
common NAT boxes and determined that UDP-hole punching works on approximately
82\% of NAT devices and TCP-hole punching on at least 64\% of NAT devices.
Unlike UDP hole punching, many TCP techniques require either super-user access
or be configured with external software packages that require super-user access.
Those that do not require super-user access require that the NAT not block
unsolicited components of the TCP three-way handshake.
The remaining devices require an intermediary or relay to pass packets
between the peers behind the NATs.  Relays or TURN style NAT traversal
presents issues similar to EtE communication in recursive routing, namely, each
node will authenticate itself with the relay, but they will not readily be
able to authenticate each other using socket-based TLS or DTLS.

\subsection{Related Works}
BitTorrent~\cite{bittorrent_security}, a P2P data sharing service,  supports
stream encryption between peers sharing files.  The purpose of BitTorrent
security is not to keep messages private but to obfuscate packets to
prevent traffic shaping due packet sniffing. Thus BitTorrent security uses a
weak stream cipher, RC4, and lacks peer authentication as symmetric keys are
exchanged through an unauthenticated Diffie-Hellman process.

Hamachi~\cite{hamachi} provides central group management and a security
infrastructure through a Web interface.  Their security system has gone through
two revisions as documented in~\cite{hamachi_security}.  Initially peers learn
of each other through Hamachi's central system, which leads to the creation of
secure links.  In their original approach, they use a system similar to a Key
Distribution Center (KDC), which requires that all security sessions initiate
through Hamachi's central servers.  In the latest version, this model has been
retained but with the addition of an external PKI, which avoids the
man-in-the-middle attack but with has the additional cost of maintaining both
an external CA and certificate revocation list (CRL).  Hamachi also supports
STUN, or NAT hole punching, and TURN style NAT traversal, though TURN requires the use of
Hamachi's own relay servers.  Because Hamachi is closed, it disables users from
hosting their own infrastructures including session management and relay
servers.

Skype~\cite{skype}, like P2PSIP, allows for decentralized audio and video
communication. Unlike P2PSIP, Skype is well-established and has millions
of users and is also closed.  While Skype does not provide documentation
detailing the security of its system, researchers~\cite{skype_auth,
skype_overview} have discovered that Skype supports both EtE and PtP security.
Though similar to Hamachi, Skype uses a KDC and does not let users setup their
own systems.

The RobotCA~\cite{robotca} provides an automated approach for decentralized
PKI.  A RobotCAs receives request via e-mail, verifies that the sender's e-mail
address and embedded PGP key match, signs the request, and mails it back to the
sender.  RobotCAs are only as secure as the underlying e-mail infrastructure
and provide no guarantees about the person beyond their ownership of an e-mail
address.  A RobotCA does not provide features to limit the signing of
certificates nor does it provide user-friendly or intuitive mechanisms for
certificate revocation.

Three approaches that propose a public overlay to create sub-overlays are
\cite{one_ring}, \cite{randpeer}, and \cite{can_multicast}.
The approach described in~\cite{one_ring} proposes the use of a universal
overlay as a discovery plane for
service overlay.  The argument is that a participant of an overlay
must support all services provided by that overlay such as multicast, DHT,
or distributed search.  Our work has the same foundations as this paper, but
takes the idea further by using the universal overlay for NAT traversal
additionally we provide mechanisms for applying PKI techniques for PtP and EtE
messaging in the service overlays.
Similarly, Randpeer~\cite{randpeer} uses a common overlay along with a
subnetting service to create individual networks for applications and services,
though the project has seen little activity and lacks implementation details.
Unlike the previous two, \cite{can_multicast} limits the sub-overlay for the
purpose of establishing multicast groups, though their approach lacks discussion
on how nodes discover and form a new overlay as such their approach is limited
to simulations.

Distributed data store applications like~\cite{dynamo, bigtable} require that
all machines have symmetric connectivity additionally like~\cite{past}, suggesting
the use of a third party application to ensure trust amongst all overlay
participants.  This is an example use case that is explicitly targeted by our
system on the presumption that there are not sufficiently easy to use
decentralized VPN software applications~\cite{sc09, nsdi10} and even if there
were it is undesirable to have additional setup requirements.

While there has been much research~\cite{secure_routing} in securing overlays
through decentralized mechanisms that attempt to prevent collusion known as
a Sybil attack~\cite{sybil}, these mechanisms do not create private overlays.
One approach mentioned does provide a natural lead into such
environments by using of a pay-to-use service to mitigate the chances
of an overlay attack, whereby the pay to use service uses a CA to sign node
IDs.  The work does not describe how to efficiently implement such a system.
Other projects~\cite{stone, tor} combine trusted overlays
with anonymous members, though it could be reasonably argued these services are
not applicable to small or medium business, which would prefer to have a private
overlay.  None of the works focus on how to apply such models to
systems that are constrained in network connectivity, e.g. by NATs.

\section{Bootstrapping Secure and Private Overlays}
\label{contributions}
In this section, we explain the individual components of our contributions.  We
begin by describing the application of SSL-derived protocols for EtE and PtP security in
overlays, which leads to how to reuse an existing public overlay to bootstrap a
private and secure overlay in a constrained environment.  We conclude with a
discussion of how we combine the components to secure group overlays through
the use of a group web interface to provide a user-friendly PKI.

\subsection{Secure Overlays}
\label{secure_overlays}
Securing EtE and PtP communication in overlays using iterative routing with
servers using static IP addresses can easily be done using TLS or DTLS with
certificates bound to the server's static IP address.  This approach does not
port well to systems that use recursive routing. First, TLS cannot easily be used
because overlay routing in NAT-constrained environments is traditionally done
through unreliable datagrams and not through reliable streams.
DTLS can be applied because it supports lost and out of order messages, though
the implementation must support usage without a socket.  The
problem then becomes how to deal with identity.  In this section, we discuss how
to bind security protocols and a certificate model to an overlay system.

\begin{figure}[h]
\centering
\includegraphics[width=3.25in]{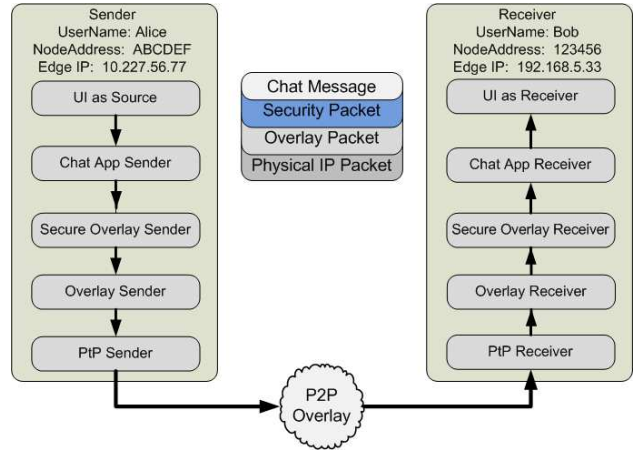}
\caption{An example of the abstraction of senders and receivers using a EtE 
secured chat application.  Each receiver and sender use the same abstracted
model and thus the chat application requires only high-level changes, such
as verifying the certificate used is Alice's and Bob's, to support security.}
\label{fig:senders_receivers}
\end{figure}

The key to our approach is abstracting the communication layer, making EtE and
PtP traffic appear identical. In this approach, all messages are datagrams, which are sent
over abstracted senders and receivers as filters illustrated in
Figure~\ref{fig:senders_receivers}.  This allows us to use secure tunnels over
these links with no application changes.

Exchanged certificates need a mechanism to verify authenticity.  Like an
Internet browser, this verification should happen automatically with no user
intervention.  Typically for SSL-derivatives, the certificates have the owner's
IP address or domain name as part of the certificate's common name field.  In
our system, we bind the certificate to each individual node ID.  That way,
a single certificate cannot be used for multiple peers in the overlay, making
it difficult for an adversary to launch Sybil attacks.

\subsubsection{Forming the Connection}
In overlay systems, a peer's connection manager requests an outgoing
connection to another peer in the system.  This triggers the creation of a
socket (UDP or TCP), which is wrapped in the abstracted sender and receiver
models.  The abstracted models arrive into the security handler, which
authenticates in both directions and creates a secure session.  The session is
wrapped in the same abstracted model and presented to the overlay system as a
direct connection to the remote peer.  To keep the system abstract, the security
model and the wrapped sockets know nothing about the overlay, and so the overlay
should verify the certificate to ensure identity.

Because EtE communication is application-specific, it requires
a slightly different path.  For that purpose, we have a specific module that
allows an application to request a secure EtE sender.
Once an application requests the sender, the
module passes a sender / receiver model to the security handler, like in the PtP
process.  Once the security initialization has completed, the
resulting sender / receiver is verified automatically for proper identity.
If that succeeds, messages sent using the EtE sender will arrive at the remote
party, decrypted and authenticated by the security handler, and delivered to the
overlay application, who will deliver to the remote party's handler for such messages.
Since overlay applications will be sending and receiving
unencrypted as well as encrypted EtE traffic, the handler must verify that the
packet was sent from a secure end point.  This assumes that an application using
an overlay has already implemented verification of node ID to some application
mapping.  For example, an application could be aware that node ID X maps to user
Y, therefore if a secure message coming from node ID Z says that it is user Y,
an application should drop the packet.

\subsubsection{Datagram Constraints}
Since UDP is connectionless, applications that use it can easily be victims of
denial of service attacks.  This is because packets sent to the receiver can
have a spoofed source address, unless the outgoing gateway prevents this from
occurring.  For each spoofed attempt, the security system will maintain state,
which can eventually be overloaded.  In TCP, this is hampered due to the
three message handshake, which verifies that a source address is not spoofed,
prior to creating security state for the connection.  To reduce the potential
of these spoofing attacks prior to establishing a secure connection, DTLS (like
Photuris~\cite{photuris}) uses a stateless cookie for each remote peer.  In
DTLS, the cookie is usually based upon the remote peer's IP and the current
time.  In our model, which deals with abstracted systems and IP addresses are
likely to be NAT-translated, this approach does not work.  Though for PtP, it
is possible to use the address and port from which the remote peer last sent
a message from.  For EtE traffic, the node ID can be used for cookie calculation.

Because we are
building on existing senders and receivers that already have state, we use the
object's memory pointer or hash value instead.  Though this leads us down a
similar path of denial of service found in TCP SYN attacks.


\subsection{Private Overlays}
\label{private_overlays}
The main components involved in the starting and maintaining a private overlay
are 1) dissemination of the security credentials and its name, 2) connecting
with and storing data in the public overlay, and 3) discovering and connecting
with peers in the private overlay.  Step 1) can be application-specific; we
propose a generic interface that is useful in many applications, through the use of groups as described in Section
Section~\ref{group_overlays}.  For 2), we presume the usage of a structured
overlay as described in Section~\ref{structured_p2p}.  In this section, we
discuss 3), the steps involved in creating and connecting to a private overlay
after the user has obtained group information and has connected to a public
overlay.

To connect with and create a private overlay, the application performs the
following steps: 1) connect to the public overlay; 2) store node ID in the
public overlay's DHT at the private group's key; 3) query the public overlay's
DHT at the private group's key; 4) start an instance of the private overlay with
the well-known end points being the node IDs retrieved from the DHT;
and 5) upon forming a link with a member in the private overlay, the node follows
the standard approach for linking to neighbors and shortcuts but using secure PtP links
to restrict connections to members of the private overlay.

The node should maintain membership in the public overlay when connected with
the private overlay.  This is needed for two reasons: first, so that other peers can discover the
node while following the same set of steps; and second, for NAT traversal purposes, as
discussed in the next paragraph.  Because the public overlay and its DHT
provides a means for discovery, nodes must maintain their node ID in the public
overlay's DHT.  A data inserter must constantly update the lease for the data
object, otherwise the data will be removed, due to the soft state or leasing
nature of DHT, whereupon (key, value) pairs are removed after a lease has expired.

\begin{figure}[h]
\centering
\includegraphics[width=2.75in]{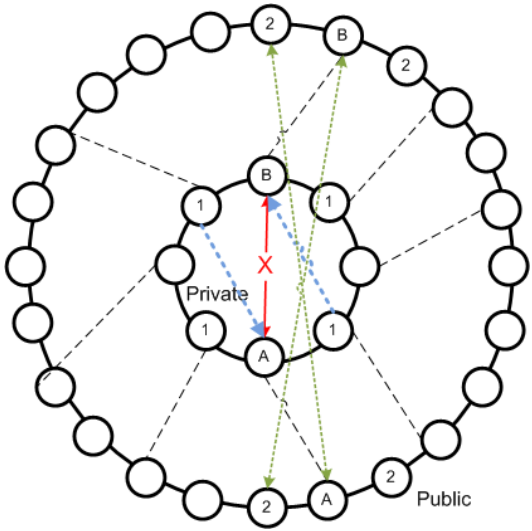}
\caption{Creating relays across the node address space, when direct
connectivity is not possible.  Two members, A and B, desire a direct connection
but are unable to directly connect, perhaps due to NATs or firewalls.  They
exchange private, 1, and public, 2, neighbor information through the private
overlay and connect to one of each other's neighbors, creating an overlap.  The
overlap then becomes a PtP relay path (represented by directed, dashed lines),
improving performance over routing across the entire overlay.}
\label{fig:overlay_relay}
\end{figure}

During the formation of the private overlay, peers may find that they are
unable to form direct connections with other members of the private overlay
even while using STUN based NAT traversal.  
We propose two solutions to address this problem: 1) to use TURN NAT traversal 
in the nodes overlay as discussed
in~\cite{nsdi10}, and 2) use the public overlay as an extra routing massive TURN
infrastructure.  The TURN NAT traversal technique has both peers connect with
each other's near neighbors in order to form a 2-hop connection with each other.
The 2-hop route can either be enforced through a static route or through EtE
greedy routing.  Due to the abstractions in the system, the public overlay can
be treated as another mechanism to create PtP links, thus while packets may use
EtE routing on the public overlay, the private overlay nodes treat it as a PtP
connection thus all communication is secured.  This approach can be further
enhanced by allowing the private overlay to apply the TURN NAT traversal
technique to the public overlay.  To do this, the private overlay must be
capable of requesting a direct connection between its node and the remote
peer in the public overlay.  This would trigger the eventual creation of a 2-hop
relay connection as presented in Figure~\ref{fig:overlay_relay}.

If overlays are small and have significant churn, it is expected that data stored
in the overlay's DHT to be lost.  This can be improved by also supporting
broadcast in the private overlay.  In this model, each peer acts as a
storage point for all data critical to itself.  If another peer cannot
successfully find data stored at a specific key in the overlay, it can make
use of broadcast over the entire overlay in an attempt to find
the result.  The technical details of our broadcast implementation are described
at the end of Section~\ref{implementation} and an evaluation of our revocation
approaches can be found in Section~\ref{evaluation_revocation}.

During our evaluation, we discovered that in certain cases the private overlay
would not form a proper well-formed state but rather more than one distinct
overlays, creating a partitioned overlay.  The underlying issue was that the
partitioned overlays believed they were in a well-formed state and thus never
reviewed the DHT list to determine if there were peers that should be their
neighbors.  This caused the overlay to remain fragmented until either a new
peer joined or enough peers left causing the nodes to believe they are in a
non-well-formed state and require bootstrapping links.
The reason the issue even existed was that states of significant churn,
especially during a bootstrapping of a significant amount of new nodes in the
system, entries in the DHT list can become partitioned, with each set of nodes potentially
seeing different lists.  Eventually the lists stored in the DHT become consistent,
but at that point, the overlay would have already been partitioned.

To proactively solve the partitioning issue, the node performs the following
steps: 1) continuously query the DHT;  2) upon receiving the DHT query result,
the node determines if there is a peer with whom it should be connected to
such as that it is closer in the address space than any of its current neighbors;
3) form a connection with that peer; and 4) the system should automatically at
this point in time realize the network fragmentation and heal itself.  In our
system, this involved creating a bootstrapping connection with the peer.  Upon
a successful connection, the system automatically causes the networks to heal.

\subsection{Group Overlays}
\label{group_overlays}
To establish trusted links, we use the PKI model, where a centralized CA (for a group) signs
all client certificates and clients can verify each other without CA interaction
by using the CA's public certificate.  However, setting up, deploying, and then
maintaining security credentials can easily become a non-negligible task,
especially for non-experts.  Most PKI-enabled systems require the use of
command-line utilities and setting up your own methods for securely
deploying certificates and policing users.  While this can be applied to an
overlay, our experience with real deployments indicates that usability is very
important, leading us to find a model with easy to user interfaces.
In this section, we present our solution, a partially automated PKI reliant on
a redistributable group based web interface.  Although this does not preclude
other methods of CA interaction, our experience has shown that it provides a
model that is satisfactory for many use cases.

\subsubsection{Joining the Group Overlay}
Membership of an overlay maps a set of users as a group. This led us to applying
the PKI model to a group infrastructure.  Using our
system, a user can host an individual or multiple groups per web site.  The
creator of the group becomes the default administrator, and users can request
access to the group.  Upon an administrator approving, users are able to
download configuration data containing overlay information and a shared key
used by the overlay application to communicate securely with the web interface.
The shared key uniquely identifies the user to the web site allowing the
application to securely send certificate requests.  By default, the web site
automatically signs all certificate requests, though it is not limited to this
model.  Two other approaches are 1) require the user to submit a request and
wait for an administrator to verify each request and 2) set a maximum amount
of automatic request signings then requiring administrative approval for more.

As stated in Section~\ref{secure_overlays}, the certificate request is bound
to the application's node ID, which can be generated by the CA or the
application.  Additionally in the group system, the certificate also contains
the user who made the request and the group for which the certificate is valid.
Not only does this ensure that a single certificate can only be used for each
node instance, but it reduces the amount of state necessary to revoke a user
from a system.  Specifically, to revoke a user, the CA would only need to
provide a signed revocation notice containing the user's name and not every one
of the previously signed certificates.

Upon receiving a signed certificate, the overlay application can connect to the
overlay where all PtP traffic will be secured and, optionally, so can EtE
traffic.  It is imperative that any operations that involve the exchanging of
secret information, such as the shared secret, be performed over a secure
transport, such as HTTPS, which can be done with no user intervention.

\subsubsection{Handling User Revocation}
Unlike decentralized systems that use shared secrets, in which the creator of
the overlay becomes powerless to control malicious users, a PKI enables the
creator to effectively remove malicious users.  The methods that we have
incorporated include:  a user revocation list hosted on the group server,
DHT events requesting notification of peer removal from the group, and
broadcasting to the entire P2P system the revocation of the peer.

A user revocation list offers an out-of-band distribution mechanism that cannot
easily be tampered, whereas communication using the overlay can be hampered
by Sybil attacks.  The revocation list is maintained on the Web site and updated
whenever an administrator removes a user, or a user leaves the group.
Additionally it can be updated periodically so that a user can verify that the revocation
list is up to date.

However, because the user revocation list requires centralization, users should
not query it prior to every communication nor periodically during conversations.
In addition to support for polling the revocation list, the use of the DHT and broadcast provides active notification of
user revocation.  Revocation through the DHT method allows a peer to request
notification if another peer is revoked from the group.  To subscribe for this
notification, the peer inserts its node ID at the peer's revocation
notification key, which we represent as a hash of its node ID.  Upon revocation,
the CA will first insert a revocation notice at this key and then query the
key for all node IDs notifying each of them of the revocation.  The insertion
of the revocation notice handles a race condition, where a peer may insert
its ID but never receive a notification.  Thus after inserting the request for
notification upon revocation, the peer should ensure that a revocation has not
occurred by querying the DHT to verify the non-existence of a CA revocation.

When the group is securing PtP traffic, the DHT approach does not effectively
seal the rogue user from the system until all peers have updated the revocation
list.  A peer may continuously connect to all peers in the system until they
have all queried the DHT key prior to verification.  Due to this issue, we 
consider an additional model in the group overlay: an overlay broadcast, ensuring
that all peers in the private overlay do know about the revocation.  

In Section~\ref{evaluation_revocation}, we present more technical details and
an evaluation with focus on latency and network traffic of the DHT and broadcast
methods.

Because the security framework is based on PKI, another approach that is also
supported is the use of certificate revocation lists
(CRL) found in most CA systems.  The advantage of a CRL and revoking individual
certificates is the ability to remove a subset of a user's node, particularly
useful in the case that the user was not malicious but that some of their nodes
had been tampered or hijacked.

\section{Applications}
\label{applications}
In this section, we present applications and potential ways to configure them
to use a private overlay.  The applications we investigate include chat rooms,
social networks, VPNs, and multicast.  The key to all these applications is that
users can easily host their own services and be discovered through the use of
a NAT-traversing, structured overlay network.

\subsection{Chat Rooms}
Chat rooms provide a platform for individuals with a common interest to find
each other, group discussion, private chat, and data exchange.  One of the most
popular chat systems for the Internet is Internet Relay Chat (IRC).  As
described in~\cite{irc}, IRC supports a distributed tree system, where clients
connect to a server, and servers use a mixture of unicast and multicast to
distribute messages.  The issues with IRC are documented by~\cite{irc_arch},
namely, scalability due to all servers needing global knowledge, reliability due
to connectivity issues between servers, and lack of privacy.  Private overlays
could be extended to support the features of IRC and potentially deal with these
inherent issues.  Each chat room would be mapped to a private overlay and the
public overlay would be used as a directory to learn about available chat rooms
and request access.  Structured overlays do not require global knowledge and can
be configured to handle connectivity issues.  Additionally, IRC by default uses
clear text messaging and even if security is used a server will be aware of the
content of the message, two issues resolved by using PtP security in a private
overlay chat room.  

\subsection{Social Networks}
Social networks such as Facebook and MySpace provide an opportunity for users to
indirectly share information with friends, family, and peers via a profile
containing personal information, status updates, and pictures.
Most social network structures rely on hosted systems, where they become
the keepers of user data, which creates privacy and trust concerns.  Private overlays
can remove this third party, making users the only owner of their data.  For this
model, we propose that each user's profile be represented by a private overlay
and that each of their friends become members of this overlay.  The overlay will
consist of a secured DHT, where only writes made by the overlay owner are valid
and only members of the overlay have access to the content stored in it.  In
addition to bootstrapping the private overlays, the public overlay would be
used as a directory for users to find and befriend each other.  For fault
tolerance and scalability, each user provides a copy of their profile
locally, which will be distributed amongst the private overlay in a read-only
DHT, therefore, allowing the user's profile to be visible whether they are
offline or online.  Each user's social network would than consist of the
accumulation of the individual private overlays and the public overlay.

\subsection{P2P VPNs}
As described in~\cite{nsdi10}, private overlays enable P2P VPNs.  The most
common type of VPNs are centralized VPNs like OpenVPN, which requires that a
centralized set of resources handle session initialization and communication.
Another approach taken by Hamachi and many others is to maintain a central
server for session initialization but allow communication to occur directly
between peers and providing a central relay when NAT traversal fails.  Using
a structured private overlay allows users to host their own VPNs, where each
VPN end points is responsible for its own session initialization and
communication.  The private overlay also provides mechanisms for handling
failed NAT traversal attempts via relaying.

\subsection{Multicast}
The topic of secure multicast has been a focus of much research.
Using an approach similar to CAN~\cite{can_multicast}, a virtual private overlay forms
a ring where all nodes are members of the multicast group with the additional
feature that you can trust that your audience is limited to those in the
overlay.  The main advantage of such multicasting technologies would be for
wide-area, distributed multicast.  Examples of such services include 
light weight multicast DNS / DNS-SD (service discovery), as well as
audio and video streaming.

\section{Implementation Details and Evaluation}
\label{evaluations}
In this section, we describe our prototype implementation of a secure, private
overlay followed by evaluation to quantify network overheads.

\subsection{Our Implementation -- Brunet}
\label{implementation}
Our implementation uses the Brunet~\cite{brunet} library, which is a P2P system
based upon the concepts introduced by Symphony~\cite{symphony}.  Its topology is a
one-dimensional ring, where peers connect with up to the four peers closest
to themselves in both directions in the node ID space.  Shortcuts are based
upon a harmonic distribution and the use of proximity~\cite{hpdc08_0}.  The
system supports distributed versions of STUN or hole-punching and TURN-like or
relay based NAT traversal~\cite{nsdi10}.  The system uses a DHT as a distributed
data store using a replication algorithm that spreads a single key, value pair
throughout the ring as described in~\cite{pcgrid07}.  Additionally, the overlay
has support for autonomic~\cite{wow} and manual creation of single-hop
connections and double-hop through relaying when NATs prevent direct
connections.  Furthermore, we have developed a P2P VPN on top of this stack,
which has been used for over three years to support a ad hoc distributed
grid~\cite{archer, gridappliance}.

The focus of this paper is on security and the bootstrapping of private pools.
In the previous sections, we have abstracted the implementation to generic
overlays, while in this section, we will present our experiences and lessons
learned in applying security protocols to the overlay.  To provide security, we
investigated two approaches:  reusing OpenSSL's DTLS implementation or making
our own platform-independent DTLS using C\# and cryptography routines provided
by .NET.  Because we are sending messages over unreliable mediums such as UDP
sockets, relays, and an overlay, we could not reuse SSL or TLS.  Additionally
we want EtE security for relays and overlay communication, the security needed
to be implemented through a filter or more specifically through a memory buffer
and not a socket.

\begin{figure}[!h]
\centering
\includegraphics[width=2.5in]{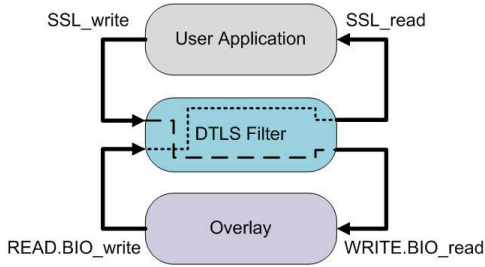}
\caption{The DTLS filter.}
\label{fig:dtls_filter}
\end{figure}

Implementation of an OpenSSL DTLS filter was non-trivial, as documentation is
sparse providing the possibility for varied approaches.  Traditionally, DTLS
uses the DGRAM (datagram) BIO (I/O abstraction) layer, which provides a reliable
UDP layer.  Because we need a filter, so that we can do both EtE and PtP traffic,
we used one memory BIO  for incoming traffic and another for outgoing traffic.
Memory BIOs provide pipes using RAM: data written to the BIO can then be read
in a first in, first out ordering.  Incoming messages written to or outgoing
messages read from the DTLS read or write BIOs, respectively, are 
either encrypted data packets or handshake control messages.  Sending and 
receiving clear text messages occur at the DTLS SSL object layer.  The pathway
for sending a clear text packet begins with the user performing an SSL\_write
operation, retrieving the encrypted data by performing a BIO\_read on the write
BIO, and sending the data over the network.  At the remote end, the packet is
passed to the SSL state machine by performing a BIO\_write on the incoming BIO
followed by a SSL\_read; the result will be the original clear text message.
This process also needs to handle control messages; we provide clear context in
Figure~\ref{fig:dtls_filter}.  As an aside, OpenSSL supports a SSL filter
BIO, though it will not work for this purpose as BIOs that are inserted are
expected to have two pipes, like a socket or two memory buffers.  Also the only
benefit of using the filter BIO would be that it manages auto-renegotiation,
which can be implemented in user code by monitoring time and received byte
count.  Other operations such as certificate verification and cookie generation
are handled by SSL callbacks, which hook into our security framework.

\begin{figure}[h]
\centering
\includegraphics[width=2.5in]{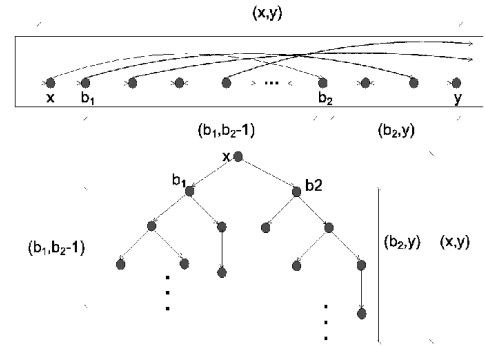}
\caption{Bounded Broadcast in range $[x, y]$}
\label{fig:tree}
\end{figure}

An important component of security is the handling of revocations handled by our
broadcasting mechanism.  We call our broadcasting model bounded-broadcast
because it is capable of broadcasting to a subset of nodes, though it is also capable
of broadcasting to the entire overlay.  A bounded broadcast uses the following
recursive algorithm:  Begin with node $x$ triggering a broadcast message over
the region $[x, y]$.  $x$ has $F$ connections to nodes in the range $[x, y]$.
Denoting the $i^{th}$ such neighbor as $b_i$, the node $x$ sends a bounded
broadcast over a sub-range, $[b_i, b_{i+1})$, to $b_i$, except the final
neighbor.  Differently stated, $b_i$ is in charge of bounded-broadcasting 
in the sub-range $[b_i, b_{i+1})$. If there is no connection to a node in the
sub-range, the recursion has ended.  The final neighbor, ($b_F$), is responsible
for continuing the bounded broadcast from $[b_F, y]$.  When a node receives a
message to a range that contains its own address the message is delivered to
that node and then routed to others in that range.  Figure \ref{fig:tree} shows
how this bounded broadcast forms a local tree recursively.   The time required
for a bounded broadcast is $O(\log^2 N)$ as shown in~\cite{small_world}.  To perform a broadcast on the entire
overlay, a peer performs the bounded-broadcast starting from its node ID with the
end address being the node ID immediately preceding its own in the address space.
Additionally for security purposes, the revoked peer is removed from the set of
neighbors, thus they are never responsible for forwarding the message onwards.

\subsection{Experimental Setup}
We use three quantitative methods to evaluate our proposed approach: network
modeling, simulation, and deployments on an actual wide-area system,
PlanetLab~\cite{planetlab}.

\subsubsection{Network Modeling}
To model our system, we implemented an application that generates a structured
overlay with the usual dimensions found in our deployed steady state systems:
a system of size N with each peer having 3 near neighbors on both sides and
$O(.5\log N)$ shortcuts.  The modeler creates a fully connected system where
shortcuts are optimally chosen based upon their location in the node ID space
using a harmonic distribution.  Once the model has been fully generated, we use
the Brunet routing code to model the number of
nodes visited and message latencies.  We employ
the MIT King data set~\cite{king_data} to determine the pair-wise latency
between nodes.  The MIT King data set consists of latency between DNS servers
distributed globally on the Internet.  Since the MIT King data set only covers
1740 nodes, for larger networks we randomly distribute the nodes in the address space placing
multiple nodes at the same ``physical'' location when necessary. The overhead due to security
was modeled by adding 3 round-trip latencies to prior to all connection
processes. This models the behavior of the 6-message DTLS handshake used in
the deployed code base.

\subsubsection{Simulations}
The approach described above estimates time based upon a model 
that reuses the core routing algorithm of the overlay code, but does not fully capture 
the dynamics of the overlay such as state machines involved in connection handling.  To allow complete
evaluation of our software stack, we have also implemented a simulator using
event-driven time that faithfully reusing the entire overlay code base including
routing, security, DHT, and connection state machines.  This allows us
to verify correct behavior in the overlay prior to testing out on real systems,
such as PlanetLab, as well as to perform experiments in a controlled
environment, which simplifies the evaluation while still retaining the effects
of a wide-are distributed system.  Like the network modeler, we have also employed
the MIT King data set~\cite{king_data} for pair wise latency between peers.
To run the simulations, we employed Archer~\cite{archer}, which uses our P2P
VPN software to form a distributed computing grid.
Because the network modeler is very
light weight, it can model more than 100,000 peers using a single computer;
however, the simulator reuses the entire overlay software and can only simulate
around 1,000 peers.

\subsubsection{PlanetLab}
PlanetLab~\cite{planetlab} is a consortium of research institutes sharing
hundreds of globally distributed network and computing resources.  PlanetLab
provides a very interesting environment as there is constant unexpected churn
of machines due to the extreme load placed on the resources and unscheduled system restarts.
Complementary to simulation, PlanetLab gives us a glimpse of what to expect
from the P2P software stack when used in an actual environment subject to higher
variance due to resource contention and churn.  Due to the time required and
complexities in working with PlanetLab, we limit the number of experiments we
used it for, focusing on the time required to join a private overlay.

\subsection{Connecting to the Private Overlay}
In this experiment, we seek to determine the overall time
it takes for a node to become connected to the public overlay, and then for a
subsequent pair node to become connected to the private overlay.  By connected, we
mean a node has a connection with the nodes whose IDs are closest in the ring (i.e., neighbors with IDs both smaller
than and larger than the node who is joining).  The results of this experiment
show the time it takes to connect to a public overlay, query the DHT for private overlay
information, and then connect to a private overlay with and without security.

To summarize the connection process, we first start the public node.  Upon
becoming connected, the private node first inserts into the DHT its 
private overlay information using an automated lease extender.  Afterwards, we
start the private node and begin constantly querying the DHT key for information
from other nodes.  As soon as this is retrieved from the DHT, it is appended to the list of
potential bootstrapping nodes from which the bootstrapping state machine pulls addresses
from during the early connection phase.  Once the private node
reaches a connected state, the test is terminated.  The reason for starting
the private node after the public node was due in part to earlier experiments,
in which we noticed that starting the public and private overlays at the same time took
longer due to the state machines in the private node using exponential time
back-offs of up to a minute when there were no nodes in the bootstrapping node list.

For PlanetLab, we created several overlays with random distribution of private
nodes.  We start from a base public overlay of
600 nodes on distinct PlanetLab hosts, then had each node randomly decide to join the private
overlay in order to obtain private overlays consisting of 5 to 600 members.  To
verify the amount of private overlay nodes, we crawled the private overlay network.
The experiment entailed connecting to the ring from the same physical location 100 times
using a different node ID each time, causing the node to join different peers and
creating a distribution of connection times. During this process, we measured the time it
took for the public node and private nodes to become connected.  To ensure that
the PlanetLab public overlay was fully connected, we waited an hour prior to starting the
experiments to provide enough time for it to reach steady state. This is a conservative
value; in our experience with overlays we run on PlanetLab, pools of this size form a complete
ring in the order of minutes.

For Simulation, we followed similar steps though we were able to tighten the
exact amount of peers in both the public and private overlays.  We began with
a pool of 700 public nodes and 2 to 300 members in the private overlay.  We
then waited for the overlays to become fully connected, whereupon we waited
60 simulated minutes to ensure that the system was at steady state, i.e.,
trimming unnecessary connections and having a steady state machine.  At which
point, we added an additional public node with a matching private node.  We
took time snapshots of how long they took to become connected.

\begin{figure}[h]
\centering
\includegraphics[width=3.25in]{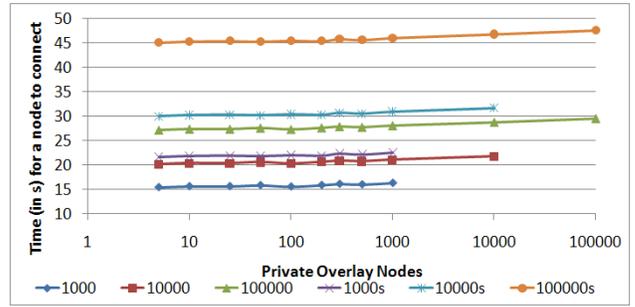}
\caption{The time it takes for a single node joining a public overlay and then
a private overlay. The X axis represents the number of peers in the private
overlay, whereas the lines themselves represent the number of peers in the
public overlay.  The legend for the line consists of ``number[s]'', where the number
represents the total numbers of peers in the public overlay and the optional ``s''
specifies whether the private overlay is modeling security.}
\label{fig:single_join_mod}
\end{figure}

For the modeler, we repeated the steps done for the simulation, though
we were able to evaluate the timing of up to 100,000 nodes, so we performed
evaluations to compare with the PlanetLab and Simulation results as well as
much larger network sizes~\ref{fig:single_join_mod}.  For timing, we reviewed
the typical connection steps when connecting to the public overlay, querying
the DHT, and then connecting to the private overlay using this process as the
basis for our evaluation.

\begin{figure}[h]
\centering
\includegraphics[width=3.25in]{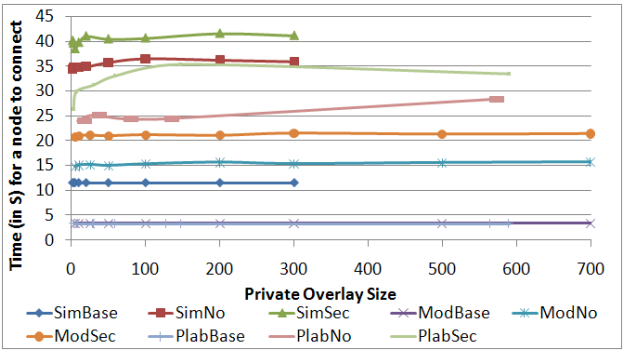}
\caption{The time it takes for a single node joining a public overlay and then
a private overlay. Since the public overlay size is the same in all tests, we
averaged all results together for the individual evaluations.  The format for
the legend is defined using the following convention: EnvironmentOverlay,
where the environment is PlanetLab, Simulator, or (Network) Modeler and
overlay is Base for public overlay, No for no security private overlay, and
Sec for security enabled private overlay. ModBase is network modeler
Public overlay connection time.  SimSec is simulator security enabled private
overlay.}
\label{fig:single_join}
\end{figure}

Our results, Figures~\ref{fig:single_join},~\ref{fig:single_join_mod} are well
correlated with near identical slopes.  In all cases the time to become
connected with the private overlay remains reasonable and scales logarithmically
as network size grows.  Connection times differ amongst the set, in the case of
the Simulator all the state machines are at the maximum wait delay due to lack of
churn in the system, whereas PlanetLab has limited overhead due to this
delay, and the modeler does not worry about state machine state,
churn, or bad connection attempts.  An important result is that PtP security
does not significantly add to time to joining the overlays, most likely due
to the majority of the time being occupied by overlay routing.

\subsection{Instantaneous Pool Creation}
\label{mass_join}
In this experiment, we determine the amount of time required to bootstrap a
private overlay including their matching public overlay nodes using an existing
network.  We start with a network size of 200 public nodes and then add the
nodes that will become part of the private overlay.  Because PlanetLab is
difficult to control in terms of attempting to start 200 public nodes
simultaneously and ensuring that they do not restart mid-experiment, we ran
this experiment using the simulator.  Though we can state from previous
experience, that using PlanetLab, we typically see a network form within
minutes of the last host's overlay software starting.

\begin{figure}[h]
\centering
\includegraphics[width=3.25in]{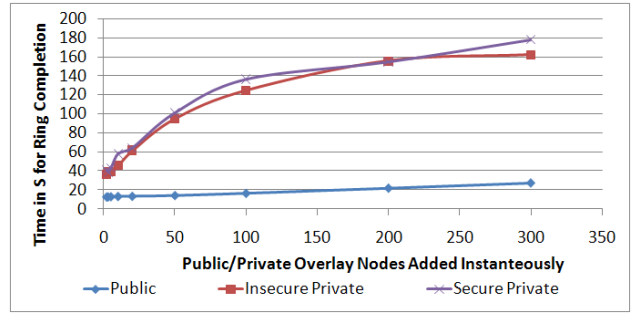}
\caption{The time to form private overlay using a public overlay with 200 nodes
after simultaneously turning on various counts of private overlay nodes.}
\label{fig:big_join}
\end{figure}

In Figure~\ref{fig:big_join}, we present our results with and without security
for various sized networks.  The results present a slightly different picture
than the previous experiment.  In this case, there is a slightly logarithmic
growth to the time it takes to complete an overlay.  As in the previous test,
the use of security has a small relative impact on the overall time to
form the ring.

\subsection{Measuring Bandwidth}
\begin{figure}[h]
\centering
\includegraphics[width=2.75in]{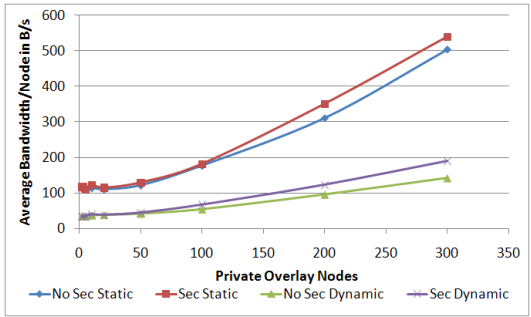}
\caption{Bandwidth used in a systems with and without security during steady state
operations consisting of 200 public nodes and various sized paired public /
private nodes.  Those lines labeled ``static'' have DHT lists queried every 5
minutes whereas in ``dynamic'' queries are made using an exponential back-off
policy starting at 30 seconds up to a maximum of 60 minutes.  Bandwidth
is in bytes / second, a negligible amount of bandwidth for DSL and Cable
Internet.}
\label{fig:bandwidth}
\end{figure}

In this experiment, we continued the overlay bootstrapped in the previous
experiment and measured the bandwidth consumed for the 60 minutes, 60 minutes 
after the overlay was well-formed.  Reusing the simulation results changes no
results and cuts down on the most expensive part of the procedure, the bootstrap
phase.  In this section, we present the network bandwidth used by each process, see
Figure~\ref{fig:bandwidth}.

As shown in Figure~\ref{fig:bandwidth}, when comparing the querying of the DHT
using static and dynamic timers, the static timer's bandwidth is dominated by
DHT queries.  Since the system is at steady state, i.e., no new nodes in the
system, only a single DHT query is made using dynamic timers.  In both cases,
the time to form a complete ring as performed in Section~\ref{mass_join} is the
same. The dynamic timer causes bandwidth to grow slower than a logarithmic pace.
Given that the bandwidth used grows slowly, it appears that overlays in general
use negligible amounts of bandwidth and that even using security does not
increase it significantly.

In regards to selecting a proper timeout, it can logically be surmised that
if an existing public and private overlay were going through heavy churn,
there will always be a base of nodes connected to each other.  If because the DHT
is currently fragmented, a new node forms a partitioned private overlay, the node
should eventually and quickly find out about the original overlay and reform
the split overlays due to the continuous queries.  The advantage of the dynamic
timeout is that it places the weight for fixing ring partitions
on the nodes that created them rather than the older more stable nodes. 

\subsection{Evaluating Revocation Implementations}
\label{evaluation_revocation}
\begin{figure}[h]
\centering
\includegraphics[width=2.75in]{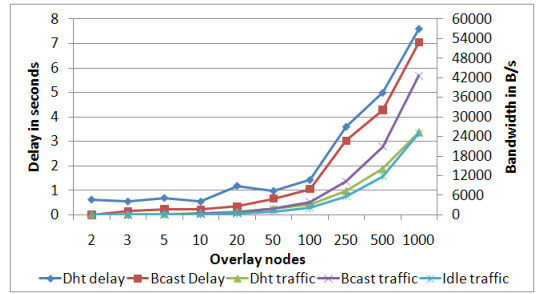}
\caption{The time delay and bandwidth used during the time between revoking a
node and notifying nodes of the revocation using simulations.  (Broadcast has
been abreviated to ``Bcast'').}
\label{fig:revocation_sim}
\end{figure}

\begin{figure}[h]
\centering
\includegraphics[width=2.75in]{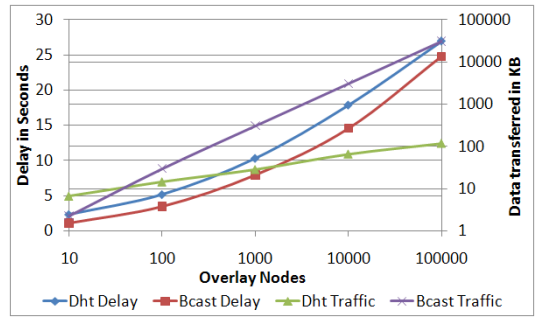}
\caption{The time delay and bytes transferred during the time between revoking
a node and notifying nodes of the revocation using simulations.  (Broadcast has
been abreviated to ``Bcast'').}
\caption{Time required and bytes transferred during a modeled DHT and broadcast
revocation.}
\label{fig:revocation_mod}
\end{figure}

In Figures~\ref{fig:revocation_sim} and \ref{fig:revocation_mod}, we present
the time required and network traffic to perform a revocation using the simulator
and modeler, respectively.  To evaluate the cost, we estimate that the average
size of a revocation is 300 bytes, which includes information such as the user's
name, the group, time of revocation, and a signature from the CA's private key.
For the simulations, we determined the average bandwidth for the 30 seconds
including the revocation as well as 30 seconds where there was no activity.
We chose this time sample, because it was the smallest amount of time that
would contain a representative steady state bandwidth but also make the
cost of a revocation obvious.  In the modeler, we only obtain total bytes
required for the operation because no connection based steady state behavior
is modeled.

The results seem to indicate that in terms of time, they scale well together,
though in network traffic the DHT scales significantly better.  The problem
with the DHT approach is that it is inefficient for truly malicious peers.
In the case of the DHT, if a peer is revoked, it can attempt to connect with
new peers who have do not know of this revocation, each peer will have to query
the DHT.  If this becomes the case, the DHT method will quickly become more
inefficient.  On the other hand, the broadcast cannot ensure that all nodes
receive the message, due to overlay network stability issues, peers may not
be included in the broadcast.  As such the best approach may be to store a
revocation in the DHT but notify all peers of a revocation via a broadcast.

Broadcast is efficient with bandwidth, because the broadcast forms a tree,
which spans exactly N-1 connections.  The network traffic required to do a
broadcast on this tree is the  minimum amount of communication necessary to
reach all nodes in the broadcast range.

%
%
%
\section{Conclusion}
\label{conclusions}
In this paper, we presented a novel architecture for deploying secure, private
overlays in constrained environments through the use of public overlays.
The public overlay at a minimum must provide a distributed data store,
like a DHT, so that peers of the private system can rendezvous with each other.
For constrained environments, we used NAT traversal techniques including STUN and
TURN with both private and public overlays to support the relaying of packets.
We evaluated the use of DTLS in our system and determined that security overhead
does not provide significant overhead during connection starting and idle periods.
Most
importantly, we presented how a group infrastructure can be used as a user-friendly
and intuitive mechanism to create and maintain private and secure overlays.  
For verification of usability, we show that the time to become connected to the
private overlay occurs in less than 30 seconds and that the bandwidth required 
from each peer while idling was insignificant, on the order of 100 bytes per
second.

At this point in time our platform still has a few drawbacks.  Users that
cannot host their own web servers on the Internet are unable to use the group's
mechanism; this could be solved by making a web site accessible via a VPN or
directly through the overlay.  For each overlay, a peer will use more
resources, to reduce this, peers could multiplex a single socket for multiple
overlays reducing the thread count and open socket count.  Also it is important
to note that the public overlay should implement decentralized security
techniques, otherwise access to the private overlay will be hampered.  For
future work, we envision applying this approach to social networks and to
establish multicast groups as done in~\cite{can}.

\bibliographystyle{IEEEtran}
\bibliography{icdcs10}

\end{document}